\documentclass[twocolumn]{aastex63}

\tightenlines

\begin{document}

\title{Formation of the cosmic-ray halo: The role of nonlinear Landau damping}


\author[0000-0003-0716-5951]{D.~O.~Chernyshov}
\affiliation{I.~E.~Tamm Theoretical Physics Division of P.~N.~Lebedev Institute of Physics, 119991 Moscow, Russia}
\author{V.~A.~Dogiel}
\affiliation{I.~E.~Tamm Theoretical Physics Division of P.~N.~Lebedev Institute of Physics, 119991 Moscow, Russia}
\author{A.~V.~Ivlev}
\affiliation{Max-Planck-Institut f\"ur extraterrestrische Physik, 85748 Garching, Germany}
\author{A.~D.~ Erlykin}
\affiliation{I.~E.~Tamm Theoretical Physics Division of P.~N.~Lebedev Institute of Physics, 119991 Moscow, Russia}
\author{A.~M.~Kiselev}
\affiliation{I.~E.~Tamm Theoretical Physics Division of P.~N.~Lebedev Institute of Physics, 119991 Moscow, Russia}

\correspondingauthor{D.~O.~Chernyshov} \email{chernyshov@lpi.ru}

\begin{abstract}
We present a nonlinear model of self-consistent Galactic halo, where the processes of cosmic ray (CR) propagation and
excitation/damping of MHD waves are included. The MHD-turbulence, which prevents CR escape from the Galaxy, is entirely
generated by the resonant streaming instability. The key mechanism controlling the halo size is the nonlinear Landau (NL)
damping, which suppresses the amplitude of MHD fluctuations and, thus, makes the halo larger. The equilibrium turbulence
spectrum is determined by a balance of CR excitation and NL damping, which sets the regions of diffusive and advective
propagation of CRs. The boundary $z_{\rm cr}(E)$ between the two regions is the halo size, which slowly increases with the
energy. For the vertical magnetic field of $\sim1~\mu$G, we estimate $z_{\rm cr}\sim 1$~kpc for GeV protons. The derived
proton spectrum is in a good agreement with observational data.
\end{abstract}




\section{Introduction}

The problem of the Galactic halo is being discussed from the beginning of the 1950s. Before that time, a sharp transition
was assumed between the Galactic disk of the thickness of $\sim100$~pc and the extragalactic medium. Later, \citet{ginz53}
developed conceptions of the physical cosmic ray (CR) halo with the size of about 15~kpc, where CRs are trapped by
scattering (i.e., propagate diffusively). The characteristic CR age in the Galaxy was estimated as $\sim 10^8$~yr, which is
confirmed by radio data and by the information on CR chemical composition (e.g., about the abundance of unstable isotope
$^{10}$Be) \citep[see, e.g.,][etc.]{Ginzburg1976,sza80}.

The static halo model of a fixed height was presented in \citet{syr64}. It assumes that the CR density at a certain distance
from the Galactic plane becomes negligible. This model is currently broadly implemented in advanced numerical codes, such as
GALPROP \citep{mosk98}.

The downside of the model is that it depends on two arbitrary parameters, namely the diffusion coefficient and the halo
size, whose values are ambiguously defined. Therefore, it is necessary to describe the processes of generation and damping
of MHD turbulence in the halo and their connections to the CR transport self-consistently.

In \citet{dogiel2020}, we have suggested a model of CR self-confinement in the Galaxy, where the turbulence generated in the
Galactic disk was amplified by streaming CRs. However, the turbulence excitation rate is very high in that model, and hence
the size of the halo is too small at GeV energies. To resolve this issue, in the present paper we also take into account the
nonlinear Landau (NL) damping, which was neglected in the original work. We show that the inclusion of the damping term
leads to a significantly larger halo size. We also show that the MHD turbulence which confines CRs in the halo can be
entirely self-generated by CRs.

\section{Self-consistent nonlinear Model of CR halo}

Unlike models with a pre-defined halo size, self-consistent halo models include a mechanism of MHD-wave excitation. In these
models, CR propagation is described by a system of nonlinear equations \citep[see, e.g., in][and references
therein]{dogiel:94,evoli18,dogiel2020}.

A general system of simplified one-dimensional nonlinear equations for the CR spectrum $N(p,z)$ and the energy density of
MHD fluctuations $W(k,z)$ can be presented in the following form:
\begin{equation}
\begin{array}{l}
{\displaystyle\frac{\partial}{\partial z} \left(u_{\rm adv}N  - D\frac{\partial N}{\partial z}\right)
-\frac{\partial}{\partial p}
\left(\frac13\frac{du_{\rm adv}}{dz}pN -\dot{p}N\right)}
= Q \,,\vspace{.2cm}\\
{\displaystyle\frac{\partial u_{\rm A}W}{\partial z} - \frac{du_{\rm A}}{dz}\frac{\partial (kW)}{\partial k}
+\frac{\partial}{\partial k}\left(\frac{kW}{\tau_{\rm cas}}\right)}=
(\Gamma_{\rm CR} - \nu)W \,,
\end{array}
\label{system}
\end{equation}
where $Q(p,z)$ is the source term of CRs, $u_{\rm adv}(z)$ is the CR advection velocity, which depends on the difference
between outward- and inward-propagating MHD waves, $u_{\rm A}(z)=B(z)/\sqrt{4\pi\rho(z)}$ is the Alfven velocity, $\dot{p} <
0$ is the rate of momentum loss due to interaction with gas, $\rho = m_p n$ is the mass density of ionized hydrogen ($m_p$
is proton mass), and $B$ is the strength of the longitudinal large-scale magnetic field. Furthermore, $\Gamma_{\rm CR}(k,z)$
is the rate of resonant wave excitation, $\nu$ is the wave damping rate, and $\tau_{\rm cas}(W)$ is the characteristic
timescale of turbulent cascade to larger $k$; the latter depends on the particular process of MHD-generation \citep[see,
e.g.,][this process is discussed in Section~\ref{sec:mirror}]{ptus06}. The spectrum $N(p,z)$ is normalized such that $\int
N(p) dp$ is the total number density of CRs.

The wavenumber $k$ of MHD fluctuations is related to the CR momentum $p$ via the resonance condition \citep{skil3},
\begin{equation}\label{resonant}
kp\approx m_p\Omega_*\,,
\end{equation}
where $\Omega_*=eB/m_pc$ is the gyrofrequency of non-relativistic CR protons. The resulting CR diffusion coefficient is
\citep{skil3}
\begin{equation}
D(p,z)\approx\frac{vB^2}{6\pi^2k^2W}\,.
\label{ddif}
\end{equation}
In this approximation, the excitation rate is proportional to the CR diffusion flux,
\begin{equation}
\Gamma_{\rm CR}(k,z)\approx-\frac{2\pi^2eu_{\rm A}p}{Bc}\,D\frac{\partial N}{\partial z}\,.
\label{gmm}
\end{equation}

There are very few known parameters and processes that can govern the density of MHD-fluctuations in the halo (and thus the
CR diffusion). These are the spatial dependencies of the magnetic field and the gas density, the magnitude and the spectrum
of CR source in the disk, and nonlinear processes involving MHD waves. In this respect, the variety of models for the wave
excitation in the halo is very restricted.

\subsection{Development of \citet{dogiel2020}} \label{sec:mirror}

\citet{evoli18} and \citet{dogiel2020} presented one-dimensional models of CR propagation along the magnetic field lines,
with MHD-fluctuations excited by the resonant CR-streaming instability.

\citet{evoli18} developed a model of MHD-turbulence with nonlinear cascading to larger $k$. They considered three sources
of waves responsible for CR scattering in the halo: (i) self-generated MHD-waves excited by CRs through the streaming
instability, (ii) processes mimicking wave generation by, e.g., supernova explosions in the disk which eject waves at large
scales, and (iii) cascading process which is determined by an initial arbitrary source of background turbulence distributed
over the halo. In case the cascading is responsible for the damping of MHD-fluctuations in the halo, the CR halo can be
about of few kpc, which is compatible with the estimations of GALPROP.

On the contrary, \citet{dogiel2020} showed that the cascading process in the halo is negligible for relevant values of $k$,
i.e., the term containing $\tau_{\rm cas}$ in the second Equation~(\ref{system}) can be omitted. We considered two sources
of waves responsible for CR scattering in the halo, namely, (i) self-generated MHD waves excited by CRs through the
streaming instability, and (ii) the spectrum of MHD-fluctuations generated by sources in the Galactic disk. In that model,
magnetic fluctuations are only excited in the direction away from the disk. However, the resulting CR flux excites waves too
efficiently, which yields the halo size of only $\sim100$~pc at low energies.

In \citet{dogiel2020}, we have not considered the possibility that outgoing waves may be reflected by a nonuniform medium
\citep[see][for details]{ferraro54,kulsrud2005}. In fact, that happens if the approximation of geometrical optics is no
longer applicable. According to \citet{Ginzburg1970}, if the wave phase velocity changes from $u_{\rm min}$ to $u_{\rm max}$
within a layer of thickness $\ell$, the reflection coefficient $R$ of the outgoing waves from the layer is
\begin{equation}
R^2 \sim \exp \left( - 4\pi k\ell\frac{u_{\rm min}}{u_{\rm max}} \right) \,.
\end{equation}
Applying this expression to the halo with $\ell = 1$ kpc, we see that even for very long waves, resonant with PeV protons,
only $\sim0.1\%$ of the total energy is reflected. Therefore, we indeed can safely assume that there are no backward-moving
waves present in the halo.

Another physical process neglected in \citet{dogiel2020} is nonlinear Landau (NL) damping. A two-dimensional halo
model including this process has already been developed by \citet{dogiel:93} and \citet{dogiel:94}. The authors
used the equation for CR propagation complemented with the equation for MHD-fluctuations which are excited by the CR flux
and attenuated by the NL damping, cascading, and adiabatic losses. It was shown that CR distribution is quasi-isotropic near
the Galactic plane, but becomes more focused along the radial coordinate as particles propagate further away. At some point
the scattering becomes unable to reflect particles back, which sets the outer boundary of the halo. The halo size was
estimated to be about 10~kpc. However the advective transport of CRs was not taken into account in this model.

According to \citet{voelk1982} and \citet{miller1991}, the rate of NL damping is given by
\begin{equation}
\nu_{\rm NL}(k) \approx g(n,T)\frac{8\pi u_{\rm A}}{B^2} k\int\limits_{k_{\rm min}}^k W(k_1) dk_1 \,,
\label{NLD}
\end{equation}
where the lower integration limit $k_{\rm min}$ is unimportant for our self-consistent model (see
Section~\ref{sec:Analytic_mid}). Following \citet{miller1991}, the dimensionless factor $g(\beta)$ in Equation~(\ref{NLD})
can be approximated by
\begin{equation}
g(\beta) \approx \frac{\sqrt{\pi}}{4}\beta^{1/2}\left(e^{-\beta^{-1}} + \frac{1}{2}\epsilon^{1/2}e^{-\epsilon\beta^{-1}}\right) \,.
\label{eq:g_approximation}
\end{equation}
Here, $\epsilon = m_e/m_p$ is electron-to-proton mass ratio and
\begin{equation}
\beta=\frac{nkT}{B^2/8\pi} \equiv \frac{u^2_{\rm th}}{u_{\rm A}^2}\,,
\end{equation}
is the plasma-$\beta$ parameter expressed via the thermal velocity of protons $u_{\rm th}$.
We assume that temperatures $T$ of both protons and electrons are equal.

The idea that the excitation of MHD turbulence in a halo can be balanced by NL damping has been previously discussed by
\citet{ptuskin97}. In the present work, we use a different expression for NL damping, which takes into account contributions
of both thermal protons and electrons. Furthermore, in our model $\beta$ significantly drops with the height, which results
in a much weaker damping for waves excited by CRs with energies above 100 GeV and, thus, helps to confine such particles.

\section{The halo model with NL damping}

\begin{figure}
\begin{center}
	\includegraphics[width=\columnwidth]{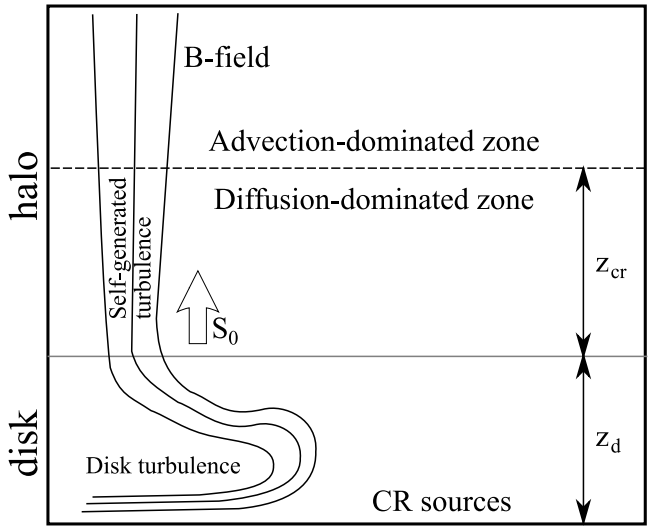}
    \caption{Schematic representation of the model considered in the paper.}
    \label{fig:schema}
\end{center}
\end{figure}

The idealized structure of our model is sketched in Figure~\ref{fig:schema}. We consider two characteristic regions along
the $z$-axis: the Galactic disk, where MHD-turbulence is assumed to be generated by sources distributed over the Galactic
plane, and the CR halo, where the turbulence is self-generated by the outgoing CR flux. We assume that the magnetic field is
practically vertical in the halo (while its geometry can be arbitrary in the disk), and that CRs do not diffuse across the
magnetic field lines.

This model allows us to reduce the set of equations~(\ref{system}) to
\begin{eqnarray}
\frac{\partial}{\partial z} \left(
u_{\rm adv}N - D\frac{\partial N}{\partial z}\right) -
\frac{\partial}{\partial p}\left(\frac{1}{3}\frac{du_{\rm adv}}{dz} pN -\dot{p}N\right)\label{eq:general_protons}\hspace{.7cm}\\
= 2Q(p)\delta(z) \,,\nonumber\\
\frac{\partial v_{\rm A}W}{\partial z} - \frac{du_{\rm A}}{dz}\frac{\partial kW}{\partial k} =
(\Gamma_{\rm CR} - \nu_{\rm NL})W \,,\hspace{2.2cm}
\label{eq:general_waves}
\end{eqnarray}
where $Q(p)$ is the source of CRs above/below the Galactic plane, $\dot{p} < 0$ is the momentum loss rate due to ionization
or proton-proton collisions within the disk ($0 < z < z_d$), while in the halo ($z\geq z_d$) the CR losses are solely due to
adiabatic cooling. Here and below, $z_d$ is the characteristic height of the disk. The CR advection velocity changes
discontinuously at the disk boundary,
\begin{equation}
u_{\rm adv}(z) = u_{\rm A}\theta(z - z_d) \,,
\label{eq:disk_to_halo_conv}
\end{equation}
where $\theta(z)$ is the Heaviside function. We assume that sources of the turbulence in the disk do not contribute to the
turbulence in the halo, i.e.
\begin{equation}
W(k, z_d) = 0 \,,
\end{equation}
i.e., the halo turbulence is entirely self-generated by CRs. However, the existence of turbulence within the disk is
essential \citep[][]{evoli18,dogiel2020}, and we take this into account in Section~\ref{sec:Analytic_strt}.

According to Equation~(\ref{eq:g_approximation}), MHD-waves are damped on plasma electrons in a low-$\beta$ plasma, and on
protons in a high-$\beta$ plasma. Most of the thermal electrons or protons contribute to the damping if, respectively,
\begin{equation}
0.01 \leq \beta \leq 0.2,
\label{range1}
\end{equation}
or
\begin{equation}
\beta \geq 10 \,.
\label{range2}
\end{equation}
For these values of $\beta$, the respective dominant exponential factor in Equation~(\ref{eq:g_approximation}) can be set to
unity, and $\nu_{\rm NL}$ in Equation~(\ref{NLD}) becomes independent of the plasma density.

Below in Section~\ref{sec:Analytic_strt} we obtain a simplified analytical solution of Equations~(\ref{eq:general_protons})
and (\ref{eq:general_waves}), while in Section~\ref{sec:num_sol} we present and discuss the exact numerical solution.

\section{Analytic approximation} \label{sec:Analytic_strt}

In this section we derive an analytical solution which qualitatively explains the role of NL damping in the self-consistent
halo model.

\subsection{CR spectrum in the disk, and CR flux from the disk to the halo}

To simplify CR propagation in the disk \citep[see, e.g.][]{ber90}, we consider the following two key parameters: the
outgoing CR flux and CR distribution function at the boundary between the disk and the halo. They both should be continuous
at boundary.

A general equation for the outgoing flux $S_0(p) = S(p,z_d)$ at the boundary is obtained by integrating
Equation~(\ref{eq:general_protons}),
\begin{equation}
S_0(p) = Q(p) + \frac{d}{dp}\left(\frac{1}{3}u_{\rm A0}\, pN_0(p)-\int\limits_0^{z_d}\dot{p}N(z,p)dz\right) \,,
\label{eq:jump_condition}
\end{equation}
where $u_{\rm A0} = u_{\rm A}(z_d)$ and $N_0(p) = N(p,z_d)$. The flux $S_0$ derived from Equation~(\ref{eq:jump_condition})
can be considered as the boundary condition for Equation~(\ref{eq:general_protons}) at $z=z_d$. Since the halo size should
be much larger than $z_d$, we assume that the CR spectrum does not change significantly across the disk. Then we can
approximate $N(p,z) \approx N_0(p)$ for $0 \leq z \leq z_d$.

As pointed out in \citet{dogiel2020}, energy losses in the halo are unimportant and thus the CR flux $S_0(p)$ is conserved.
Then we obtain the following solution of Equation~(\ref{eq:general_protons}) for $z \geq z_d$:
\begin{equation}
N(p,z) = \frac{S_0(p)}{u(p,z)}\,,
\label{eq:f_general_solution}
\end{equation}
where $u(p,z)$ is the outflow velocity of CRs,
\begin{equation}
u = \left(\int\limits_\eta^{\eta_\infty} \frac{e^{\eta-\eta_1}d\eta_1}{u_{\rm A}(\eta_1)} \right)^{-1} \,,
\end{equation}
and $\eta(p,z)$ is a dimensionless variable
\begin{equation}
\eta = \int\limits_{z_d}^z \frac{u_{\rm A}}{D} dz_1 \,.
\label{eq:eta}
\end{equation}
The value of $\eta_\infty$ generally depends on the boundary condition at $z \rightarrow \infty$. Substituting $N_0(p)=N(p,
z_d)$ from Equation~(\ref{eq:f_general_solution}) in Equation~(\ref{eq:jump_condition}), we derive the flux,
\begin{equation}
S_0(p) = \frac{u_d(p)}{\mathcal{E}(p)}\int\limits_p^\infty Q(p_1) \exp \left( - \int\limits_p^{p_1}
\frac{u_d(p_2)}{\mathcal{E}(p_2)} dp_2 \right) dp_1 \,.
\label{eq:jump_expression}
\end{equation}
Here, $u_d(p) = u(p,z_d)$ and $\mathcal{E}(p) = \frac13pu_{\rm A0}-\int_0^{z_d}\dot{p}dz = \frac13pu_{\rm A0} +
\frac12\mathcal{N}_{\rm H} L(p)$, where $\mathcal{N}_{\rm H}$ is the vertical column density of hydrogen atoms in the disk
and $L(p) = -\dot{p}/n_{\rm H}$ is energy loss function (per unit column density) due to interaction with the disk gas. We
note that $u_d$ in fact depends on $S_0$, and, therefore, Equation~(\ref{eq:jump_expression}) is an integral equation for
$S_0(p)$. If the dependence $u_d$ versus $S_0$ is weak, the equation can be solved
iteratively. 

We can obtain a simple approximation for $S_0(p)$ assuming that $\mathcal{E}S_0/u_d$ is a power-law function $\propto
p^{-\alpha}$. According to experimental data, $S_0/u_d\equiv N(p)$ has a negative spectral index smaller than that of
$S_0(p)$ (both are smaller than $-2$), while $\mathcal{E}$ cannot increase faster than $\propto p$. Therefore, $\alpha>0$
and we readily obtain from Equation~(\ref{eq:jump_condition}),
\begin{equation}
S_0(p) = \frac{Q(p)}{1 + \alpha\mathcal{E}/(pu_d)} \,.
\end{equation}
We notice that both the energy loss rate $\mathcal{E}/p$ and the inverse outflow velocity $u_d^{-1}=N/S_0$ decrease with
$p$, and therefore $S_0(p) \approx Q(p)$ for sufficiently high energies. To evaluate a critical energy where
$\mathcal{E}/(pu_d)=1$, we assume $\mathcal{N}_{\rm H} \approx 6\times 10^{20}$~cm$^{-2}$ and $u_d \sim u_{\rm A}\sim
10^6$~cm/s. This yields the proton energy about 0.5~GeV, above which we can set $S_0 = Q$.

As discussed in \citet{dogiel2020}, the value of $\eta$ is the key parameter characterizing CR propagation in the halo. The
entire halo can be approximately split into two regions: one is called the {\it halo sheath}, where $\eta(z) \ll 1$ and the
diffusion term $-D\,\partial N/\partial z$ dominates in the CR flux $S_0$; the other is where $\eta(z) \gg 1$ and the
advection term $u_{\rm adv}N$ dominates. The critical point $z_{\rm cr}(p)$ separating these two regions can be determined
from the condition $\eta(p,z_{\rm cr})\approx1$. Since the dominant advection irreversibly carries CRs away from the disk,
the halo size can be set equal to $z_{\rm cr}$. Therefore, the boundary condition at $z \rightarrow \infty$ becomes
unimportant as long as $\eta_\infty \gg 1$.

In order to derive $\eta$ from Equation~(\ref{eq:eta}), we need to obtain the diffusion coefficient $D$ from
Equation~(\ref{ddif}), which requires the solution of Equation~(\ref{eq:general_waves}).

\subsection{Excitation-damping balance}\label{sec:Analytic_mid}

The numerical solution of Equations~(\ref{eq:general_protons}) and (\ref{eq:general_waves}) (see
Section~\ref{sec:num_sol}) suggests that $W(k,z)$ in the diffusion region can be estimated from the excitation-damping
balance,
\begin{equation}
\Gamma_{\rm CR} = \nu_{\rm NL}\,.
\end{equation}
We rewrite it using Equations~(\ref{gmm}) and (\ref{NLD}),
\begin{equation}
\frac{4g(z)c^2}{\pi e^2B^2} k^2\int \limits_{k_{\rm min}}^k W(k_1) dk_1 = S_0(p) - u_{\rm A}N\,.
\label{gz}
\end{equation}
In the halo sheath ($\eta < 1$) the last term of Equation~(\ref{gz}) can be neglected. In this case, $S_0(p)\propto Q(p)$
decreases with $p$ faster than $p^{-2}$, and thus the integral on the left-hand size of Equation~(\ref{gz}) is dominated by
the upper limit $k$. Therefore,
\begin{equation}
W(k,z) = \frac{\pi}{4g(z)}\frac{\partial}{\partial k}\left[p^2S_0(p)\right] \,,
\end{equation}
and
\begin{equation}
\eta(p,z) = -\frac{3\pi^3e}{2vc}\frac{\partial}{\partial p}\left[p^2S_0(p)\right] \int\limits_{z_d}^{z}
\frac{u_{\rm A}}{Bg(z_1)}dz_1 \,.
\end{equation}
From  Equation~(\ref{eq:g_approximation}) we obtain
\begin{equation}
\eta = -\frac{6\pi^{5/2}e}{vc}\frac{\partial}{\partial p}\left[p^2S_0(p)\right] \int\limits_{z_d}^{z}
\frac{u_{\rm th}(T)\beta^{-1}dz_1}{B(e^{-\beta^{-1}} + \frac12\epsilon^{1/2}e^{-\epsilon\beta^{-1}})} \,.
\label{eq:eta_notexpanded}
\end{equation}
For simplicity, below we assume $n(z)=n_0 \exp(-z/z_n)$, $B(z)=B_0 \exp(-z/z_B)$, and $T(z)=T_0 \exp(z/z_T)$.

\subsection{Spectrum of CRs in the halo sheath} 
\label{sec:Analytic_end}

If $\beta$ is within the ranges defined in Equations~(\ref{range1}) and (\ref{range2}), the expression for $g(\beta)$
simplifies significantly. In this regime, previously considered by \citet{ptuskin97},  we can neglect the exponential
dependence in the denominator of Equation~(\ref{eq:eta_notexpanded}) and rewrite the equation as
\begin{equation}
\eta(p,z) = A(p) \left(e^{z/z_\eta} - e^{z_d/z_\eta}\right)\,,
\label{eq:eta_const_damping}
\end{equation}
where $z_\eta^{-1} = z_n^{-1} - z_B^{-1} - \frac12z_T^{-1}$. For the sake of simplicity, below we assume $z_d \approx 0$.

The magnitude of the dimensionless factor $A(p)$ depends on the dominant mechanism of NL damping. In a low-$\beta$ plasma
with $0.01 \leq \beta \leq 0.2$ the damping on thermal electrons dominates, and
\begin{equation}
A(p) \approx A_e(p) = -\frac{12\pi^{5/2}eu_{\rm A0}^2z_\eta}{vcB_0u_{\rm th0}\epsilon^{1/2}}\,\frac{\partial}{\partial p}
\left[p^2S_0(p)\right] \,,
\label{eq:alow_beta}
\end{equation}
where $u_{\rm th0} = u_{\rm th}(z_d)$, while in a high-$\beta$ plasma with $\beta \geq 10$ the damping is due to thermal
protons, and
\begin{equation}
A(p) \approx A_p(p) =  \frac12\epsilon^{1/2}A_e(p)\,.
\label{ahp}
\end{equation}
The critical point $z_{\rm cr}$ is derived from the condition $\eta \approx 1$. Thus, the halo size is estimated from
Equation~(\ref{eq:eta_const_damping}) as
\begin{equation}
z_{\rm cr}(p) = z_\eta \ln \left[1 + 1/A(p) \right] \,.
\label{z_cr}
\end{equation}
For low energies, where $A(p) \gg 1$, the halo size increases with $p$ as $z_{\rm cr}(p)\approx z_\eta/A(p)$; for high
energies, the halo size $z_{\rm cr}(p) \approx -z_\eta \ln A(p)$ is almost independent of $p$. We point out that the
model is not viable in the former regime, normally corresponding to the electron-dominated damping, because the resulting
halo size becomes too small. On the other hand, for the proton-dominated damping with $\beta > 10$, the function $A_p(p)\sim
1$ for the following halo parameters:  $B \leq 1~\mu$G, $n \geq 10^{-2}$~cm$^{-3}$, and $T \geq 100$~eV. The halo size in
this case exceeds 1~kpc at energies above 1~GeV.

The CR spectrum is given by Equation~(\ref{eq:f_general_solution}),
\begin{equation}
N(p,z) = \frac{S_0(p)}{u_{\rm A0}}\int\limits_\eta^\infty \frac{e^{\eta-\eta_1}d\eta_1}{[1 + \eta_1/A(p)]^{z_\eta / z_{\rm A}}} \,,
\end{equation}
where $z_{\rm A}^{-1} = \frac12z_n^{-1} - z_B^{-1}$ characterizes the spatial scale of variation of $v_{\rm A}(z)$. This
result can be expressed in terms of the incomplete gamma-function $\mathbf{\Gamma}(a,z)$,
\begin{eqnarray}
N(p,z) = \frac{S_0(p)}{u_{\rm A0}}e^{\eta + A(p)}A(p)^{z_\eta/z_{\rm A}}\hspace{2cm} \\
\times\mathbf{\Gamma}(1 - z_\eta/z_{\rm A}, A(p) + \eta) \,.\nonumber
\end{eqnarray}
If $A(p) + \eta \gg 1$, the solution corresponds to the advection flux with the Alfven velocity at the halo periphery. This
represents low-energy CRs at large distances from the disk,
\begin{eqnarray}
N(p,z) \approx \frac{S_0(p)}{u_{\rm A0}}[1 + \eta/A(p)]^{-z_\eta / z_{\rm A}}\hspace{2cm} \label{eq:f_const_low}\\
= \frac{S_0(p)}{u_{\rm A0}}e^{-z / z_{\rm A}}\equiv \frac{S_0(p)}{u_{\rm A}(z)} \,.\nonumber
\end{eqnarray}
If $A(p) + \eta \ll 1$, the CR spectrum tends to
\begin{eqnarray}
N(p,z) \approx \frac{S_0(p)}{u_{\rm A0}} A(p)^{z_\eta/z_{\rm A}} \mathbf{\Gamma}(1 - z_\eta/z_{\rm A})\hspace{1.6cm}
\label{eq:f_diffusion_appx}\\
\propto S_0(p)[pS_0(p)]^{z_\eta/z_{\rm A}}\,.\nonumber
\end{eqnarray}
The resulting spectrum, corresponding to the diffusion-dominated flux, does not practically depend on $z$ up to the critical
point $z_{\rm cr}$.  We note that Equation (\ref{eq:f_diffusion_appx}) resembles Equation~(34) from \citet{ptuskin97}.

The derived approximate solution has important implications. We conclude that CRs escape from the halo at $z=1-10$~kpc, and
for energetic CRs the halo size weakly depends on their energy. Given $S_0(p) \propto p^{-2.4}$ and $N(p) \propto p^{-2.7}$,
our solution suggests that $z_\eta/z_{\rm A} \approx 0.3/1.4$ or $z_n/z_B - 0.15z_n/z_T = 0.35$.

\section{Numerical solution and discussion}\label{sec:num_sol}

Equations~(\ref{eq:f_const_low}) and (\ref{eq:f_diffusion_appx}) provide sufficiently good approximations for the CR
spectrum as long as $\beta$ is within the ranges defined in Equations~(\ref{range1}) and (\ref{range2}). However, the
magnitude of $\beta$ varies strongly with $z$ and, therefore, the exponential terms in the denominator of
Equation~(\ref{eq:eta_notexpanded}) cannot be generally ignored. As a result, the expressions for $\eta$ and $N$ become
complicated and can only be obtained numerically.

To reduce the number of free parameters, we consider a simple isothermal model ($z_T^{-1}=0$) with a constant magnetic field
($z_B^{-1} =0$). We use the following set of parameters: $B = 1~\mu$G, $n_0 = 0.1$~cm$^{-3}$, $z_n = 1$~kpc, and $T =
10$~eV. For the CR source function, we use $Q(p) \simeq Q_* (p/m_pc)^{-2.4}$ with $Q_*m_pc = 9.4\times
10^{-4}$~cm$^{-2}$~s$^{-1}$, which is similar to the value given by \citet{strong10}. In this case, $\beta = 40$ at $z = 0$.
The total power of CR sources in the Galaxy can be roughly estimated as $\mathcal{W} = 2\pi R_{\rm Gal}^2\int E_{\rm
kin}(p)Q(p) dp = 4.5\times 10^{-3}~\mbox{erg}~\times 2\pi R_{\rm Gal}^2Q_*m_pc$. Assuming $R_{\rm Gal} = 20$ kpc for the
Galactic disk radius, we obtain $\mathcal{W} \approx 10^{41}$ erg/s. 

Equations~(\ref{eq:general_protons}) and (\ref{eq:general_waves}) are solved numerically by employing the procedure
described in \citet{dogiel2020}. To account for CR species heavier than protons, the excitation rate $\Gamma_{\rm CR}$ is
multiplied by a factor of 1.5 \citep[see, e.g.,][]{dogiel2018}. For the initial CR density we use $N(p,z) = 0$, to avoid
appearance of a sharp discontinuity at the upper halo boundary. Both the initial MHD spectrum and the boundary condition at $z = 0$
are equal to a small non-zero function $W_0(k,z)$, as it is necessary for the waves excitation. To ensure a weak (logarithmic)
dependence of the integral in Equation~(\ref{NLD}) on the integration limits, we use $W_0(k,z) \propto k^{-1}$.

\begin{figure}
\begin{center}
	\includegraphics[width=\columnwidth]{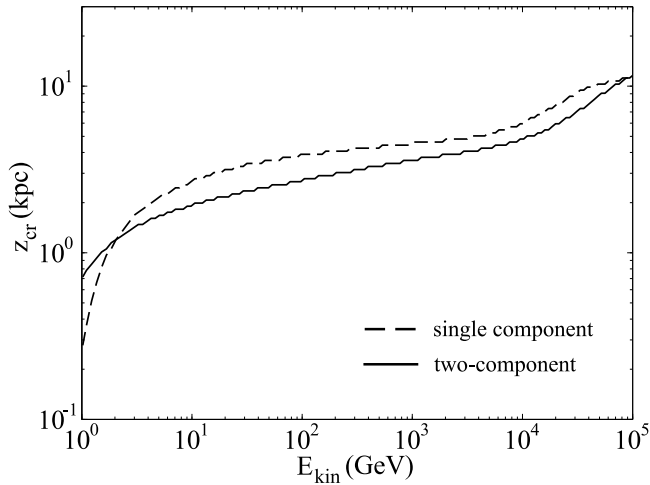}
    \caption{Halo size $z_{\rm cr}(E_{\rm kin})$ obtained from the numerical solution of our model. The dashed line shows the case of
    a single-component gas with $B = 1~\mu$G, $n_0 = 0.1$~cm$^{-3}$, $z_n = 1$~kpc, and $T = 10$~eV ($\beta = 40$),
    the solid line represents the case of a two-component gas (see Section~\ref{sec:num_sol}).}
    \label{fig:size_halo_comparison}
\end{center}
\end{figure}

\begin{figure}
\begin{center}
	\includegraphics[width=\columnwidth]{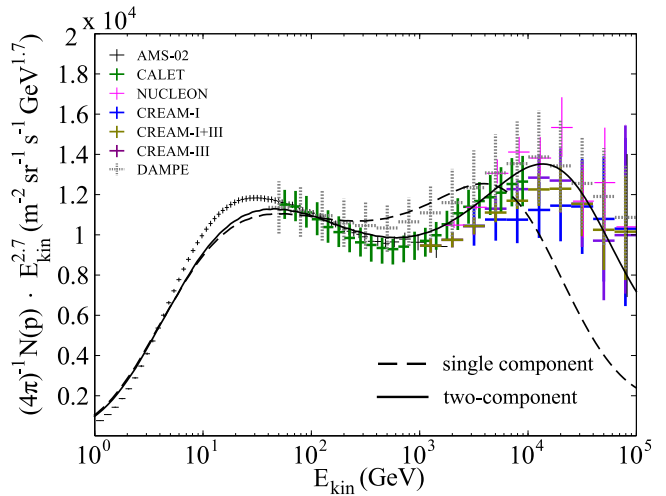}
    \caption{Energy spectra of CR protons obtained from the numerical solution of our model (lines) and the observational data
    (symbols). All parameters are the same as in Figure~\ref{fig:size_halo_comparison}.}
    \label{fig:CR_spectra}
\end{center}
\end{figure}

The resulting halo size, $z_{\rm cr}$, and the differential CR spectrum, $N/4\pi$, are plotted versus the proton kinetic
energy $E_{\rm kin}$ by the dashed lines in Figures~\ref{fig:size_halo_comparison} and \ref{fig:CR_spectra},
respectively. To account for the solar modulation, we use the force-field approximation with potential $\phi = 0.5$ GV
\citep[][]{gleeson68}. Observational data are taken from \citet{agu15} (AMS-02), \citet{calet19} (CALET), \citet{nucleon19}
(NUCLEON), \citet{cream11} (CREAM-I), \citet{cream17} (CREAM-I+III), and \citet{an2019} (DAMPE). The data are collected
using Cosmic-Ray DataBase (CRDB v4.0) by \citet{Maurin2020}.

We stress that the CR spectra strongly depend on a particular model of NL damping. In our case, the damping is
described by Equations~(\ref{NLD}) and (\ref{eq:g_approximation}). Since $\beta$ rapidly drops with the height, so does the
damping and, hence, the CR diffusion coefficient. Therefore the CR spectra plotted in Figure~\ref{fig:CR_spectra} can be
interpreted as follows: 
\begin{itemize}
\item $E_{\rm kin} < 10$ GeV: At such energies, $A(p)$ is sufficiently large and, thus, the halo size is small in
    accordance with Equation~(\ref{z_cr}). For this reason, $\beta(z_{\rm cr})\approx \beta(0)>10$ and NL damping is due
    to thermal protons. Equation~(\ref{eq:f_const_low}) is applicable, which gives $N(p) \propto Q(p) \propto p^{-2.4}$.
\item $10~\mbox{GeV} < E_{\rm kin} < 1~\mbox{TeV}$: $N(p)$ starts approaching a softer spectrum described by
    Equation~(\ref{eq:f_diffusion_appx}). The halo size increases with energy as $z_{\rm cr}(p)\propto 1/A(p)$, and thus
    $\beta(z_{\rm cr})$ rapidly decreases, so that eventually a mixed damping both on thermal protons and electrons
    operates.
\item $100~\mbox{GeV} < E_{\rm kin} < 10~\mbox{TeV}$: In the mixed-damping regime, a smooth transition from $A_p(p)$ to
    much larger $A_e(p)$ occurs. According to Equation~(\ref{eq:f_diffusion_appx}), that makes $N(p)$ harder
		(NL damping rapidly reduces with CR energy as the proton contribution becomes negligible,
		and therefore the CR confinement increases). In Figure~\ref{fig:CR_spectra}, the transition is manifested by
the increase seen at $1~\mbox{TeV} < E_{\rm kin} < 10~\mbox{TeV}$.
\item $E_{\rm kin} > 10~\mbox{TeV}$: Finally, at very high energies $\beta(z_{\rm cr})$ decreases below $0.1$, where the
    damping is due to thermal electrons. Equation~(\ref{eq:f_diffusion_appx}) becomes applicable; since $z_\eta / z_{\rm
    A} = 1/2$ in our case, $N(p) \propto p^{-3.1}$.
\end{itemize}

Figure~\ref{fig:CR_spectra} shows that the theoretical curve and the experimental data are in good qualitative agreement.
However, we should also keep in mind that gas in the halo consists of several components. In particular, the warm ionized
gas (WIM) dominates at lower altitudes, while at higher $z$ it is mostly hot coronal gas \citep{ferriere1998,gaen08}. To
account for multiple gas components, we assume that the total gas density in our model is determined by a sum of the two
phases: $n(z) = n_{\rm hot}(z) + n_{\rm WIM}(z)$. The same principle applies to the magnitude of NL damping in
Equation~(\ref{NLD}): $g(z) = g(\beta_{\rm hot}) + g(\beta_{\rm WIM})$. Note that the factor $ku_{\rm A}$ in
Equation~(\ref{NLD}) is the wave frequency, and therefore is the same in both phases. Assuming $B = 1~\mu$G, we use the
following set of parameters:
\begin{itemize}
\item Warm phase ($\beta = 4$): $n_0 = 0.1$~cm$^{-3}$, $T = 1$~eV, $z_n = 0.4$~kpc.
\item Hot phase ($\beta = 4$): $n_0 = 10^{-3}$~cm$^{-3}$, $T = 100$~eV, $z_n = 2$~kpc.
\end{itemize}
The source function is $Q(p) \simeq Q_*(p/m_pc)^{-2.32}$ with $Q_*m_pc = 9\times 10^{-4}$~cm$^{-2}$~s$^{-1}$.

The results for the two-phase model are depicted in Figures~\ref{fig:size_halo_comparison} and \ref{fig:CR_spectra} by the
solid lines. We see that the theoretical curve show a much better agreement with the observational data in this case, which
is due to a much weaker dependence of $\beta$ on $z$.

While the two-phase model provides a remarkably good overall agreement with the experimental data in a wide energy range,
the discrepancy below 10 GeV is up to $20\%$. We believe that this is because the effect of disk turbulence on the vertical
profile of the spectrum can no longer be ignored at such low energies. Indeed, by deriving
Equation~(\ref{eq:jump_expression}) we assume that $N(p,z_d) = N(p,0)$. While this assumption is certainly reasonable for
high energies, the low-energy part of the spectrum should be stronger affected by the fact that the diffusion coefficient in
the disk decreases with energy, which inevitably leads to an increasing vertical gradient of $N(z)$. Therefore, the
low-energy spectrum should be more inhomogeneous at $0 < z < z_d$, and the actual spectrum at $z = 0$ should go somewhat
above the theoretical curves plotted in Figure~\ref{fig:CR_spectra}. Furthermore, the diffusion coefficient in the
Galactic disk is likely not affected by the CR streaming (e.g., due to heavy damping on neutrals), but rather depends on
external sources of turbulence (such as supernova explosions and stellar winds).

Apart from the halo size, another important parameter characterizing propagation of CRs is their grammage $X$, i.e.,
the average surface density traversed by CRs during their lifetime in the Galaxy. The grammage determines the ratio of
secondary-to-primary nuclei, and thus can be derived from experimental data. For our model, it can be roughly estimated as
\begin{equation}
X \approx \mathcal{N}_{\rm H}m_p \frac{c}{u_d} \,,
\end{equation}
which gives $X(10~\mbox{GeV}) \approx 12$ g/cm$^2$ for our parameters at $E_{\rm kin} = 10$~GeV. This value is close to that
obtained by, e.g., \citet{engelmann90}. We stress, however, that such estimates are very approximate: to properly test the
model, we need to accurately calculate the spectra of secondary and primary nuclei, and compare them to the experimental
data. This work will be reported in a separate paper. 

\section{Conclusions}

We present a development of the self-consistent model of the Galactic CR halo, extending the model by \citet{dogiel2020}.
Our earlier model by \citet{dogiel2020} predicts a small size of the halo at low energies, which does not agree with
experimental data. To overcome this discrepancy, we include nonlinear Landau (NL) damping in the present model.

The key input parameters of the proposed model are the CR source $Q(p)$ as well as the spatial profiles of the vertical
magnetic field $B$, ionized gas density $n$, and temperature $T$. We show that all these parameters may significantly affect
the size of the halo, in particular at relatively low CR energies. The MHD-turbulence in the halo, which controls the
vertical escape of CRs, is entirely generated by the resonant CR-streaming instability. The equilibrium spectrum of MHD
waves in our present model is reached when the CR excitation rate is balanced by NL damping. This significantly suppresses
the amplitude of MHD waves compared to the model of \citet{dogiel2020}, thus making the halo size substantially larger.

We consider two alternative models of gas distributions in the halo: a single-component isothermal model and a two-phase
model composed of hot coronal gas and warm ionized gas. We showed that the single-component model requires very dense and
hot gas with $\beta \approx 40$ at low altitudes to be able to reproduce the experimental data. For the two-phase model, the
required gas parameters are much closer to those reported in the literature \citep[e.g.,][]{ferriere1998}.

Our model is able to reproduce the spectrum of CR protons in a wide range of energies, including the spectral features
observed between $\sim10$~GeV and $\sim10$~TeV (see Fig. \ref{fig:CR_spectra}). Despite some 20\% discrepancy with
experimental data below 10~GeV, our model predicts a reasonable halo size of about 1~kpc at 1~GeV. We argue that such a
discrepancy may be due to increasing influence of the Galactic disk at lower energies, which is still neglected in our
model.

\acknowledgments

The authors are grateful to an anonymous referee for constructive suggestions, and to Andrey Bykov for useful
discussions and comments. The work of DOC, VAD, ADE, and AMK is supported by the Russian Science Foundation via the Project
20-12-00047.

\section*{Note added in proof}
New data on CR proton spectrum reported by CALET \citep{adriani22} confirms the existence of the second spectral break at 10 TeV. The break position suggests that the scale height of hot gas should be about 2 kpc or less.

\bibliographystyle{apj}
\bibliography{halo_NLD_bio}

\end{document}